\documentclass[preprint]{aastex62} 

\usepackage{xspace}
\usepackage{bm}

\newcommand{\D}{\mathrm{d}}
\newcommand{\checkme}[1]{{#1}}
\newcommand{\highspinodds}{\checkme{15:1}\xspace}
\newcommand{\highspinsigma}{\checkme{$1.8 \sigma$}\xspace}
\newcommand{\alignedodds}{\checkme{22000:1}\xspace}
\newcommand{\alignedsigma}{\checkme{$4 \sigma$}\xspace}

\graphicspath{{./}{figures/}}

\submitjournal{ApJ}

\begin{document}

\title{Constraining black-hole spins with gravitational wave observations}

\correspondingauthor{Vaibhav Tiwari}
\email{tiwariv@cardiff.ac.uk}

\author{Vaibhav Tiwari}
\affil{Cardiff School of Physics and Astronomy \\
Cardiff University, Queens Buildings, The Parade \\
Cardiff CF24 3AA, UK}

\author{Stephen Fairhurst}
\affil{Cardiff School of Physics and Astronomy \\
Cardiff University, Queens Buildings, The Parade \\
Cardiff CF24 3AA, UK}

\author{Mark Hannam}
\affil{Cardiff School of Physics and Astronomy \\
Cardiff University, Queens Buildings, The Parade \\
Cardiff CF24 3AA, UK}



\begin{abstract}

The observation of gravitational-wave signals from merging black-hole binaries enables direct measurement of the properties of the black holes. An individual observation allows measurement of the black-hole masses, but only limited information about either the magnitude or orientation of the black hole spins is available, primarily due to the degeneracy between measurements of spin and binary mass ratio. Using the first six black-hole merger observations, we are able to constrain the distribution of black-hole spins.  We perform model selection between a set of models with different spin population models combined with a power-law mass distribution to make inferences about the spin distribution.  We assume a fixed power-law mass distribution on the black holes, which is supported by the data and provides a realistic distribution of binary mass-ratio.  This allows us to accurately account for selection effects due to variations in the signal amplitude with spin magnitude, and provides an improved inference on the spin distribution. We conclude that the first six LIGO and Virgo observations \citep{O1:BBH, gw170104, GW170608, GW170814} disfavour highly spinning black holes against low spins by an odds-ratio of \highspinodds; thus providing strong constraints on spin magnitudes from gravitational-wave observations. Furthermore, we are able to rule out a population of binaries with completely aligned spins, even when the spins of the individual black holes are low, at an odds ratio of \alignedodds, significantly strengthening earlier evidence against aligned spins \citep{Farr_2017}. These results provide important information that will aid in our understanding on the formation processes of black-holes.

\end{abstract}

\keywords{black hole spins, gravitational waves}

\section{Introduction} \label{sec:intro}
Gravitational waves (GW) emitted by merging black holes are identified in the LIGO and Virgo data through the use of 
search analysis pipelines, which use the known waveform morphology to identify weak signals in the 
data \citep{O1:BBH, gw170104, GW170608, GW170814}.  The observations are followed by parameter-estimation analyses that extract posterior probability distributions for the parameters of the binary --- the masses and spins of the component black holes as well as the distance, sky location and orientation of the binary \citep{Cutler:1994ys, first_monday_pe, Veitch:2014wba}. While some parameters are extracted with good precision, others cannot be accurately measured, and several sets of parameters are strongly correlated, for example the distance with binary orientation, and mass ratio with black hole spins.
Nonetheless, the observed parameters from several observations can be combined to obtain the underlying astrophysical distributions of black-hole masses and spins.  In this paper we use publicly available information of the measurements from the first six GW signals observed from merging black holes (GW150914, LVT151012, GW151226 and GW170104, GW170608 and GW170814)\citep{O1:BBH, gw170104, GW170608, GW170814} to draw inferences about the underlying spin distribution of black holes \footnote{We generate parameter distribution using confidence intervals reported in observation papers -- please see section on method used in this paper.}.

\subsection{Model Selection}
For much of the parameter space it is not possible to accurately measure the individual spins with LIGO-Virgo observations at typical SNRs \citep{Purrer:2015nkh}, but only two mass-weighted combinations of the  spins: an effective spin, $\chi_{\mathrm{eff}}$, which describes the effect the aligned components have on the inspiral rate at which the binary's orbit decays and the maximum orbital frequency prior to merger \citep{Ajith:2009bn}, and an effective precessing spin, $\chi_p$, which determines the effects of precession and leads to modulations in the observed waveform \citep{acst:1994, Schmidt:2014iyl}. For the reported observations, there has been no discerning evidence for precession (the posterior distributions for $\chi_{p}$ are consistent with the priors). Consequently, we restrict attention to the observed values of $\chi_{\mathrm{eff}}$.

The effective spin used in LIGO-Virgo analyses is related to the individual spins of the two black holes in the binary by~\citep{Damour:2001, Ajith:2009bn}
\begin{equation}\label{eq:chi_eff}
 \chi_{\mathrm{eff}} = \frac{m_1\chi_1 + m_2\chi_2}{m_1 + m_2},
\end{equation}
where $m_1$ and $m_2$ are the component masses of the binary and $\chi_1$ and $\chi_2$ are the components of the (dimensionless) spins aligned with the angular momentum defined as $\chi = c\mathbf{S\cdot \hat{L}}/(Gm^2) \leq 1$. 

We are interested in inferring the spin distribution of the merging black holes binaries in the Universe by comparing the observed distribution of effective spins with those predicted by astrophysical models. At present, we have only a limited number of observations and therefore we focus on a discrete set of models. We follow \citep{Farr_2017} and introduce six possible spin distributions for comparison.  We use three distributions for spin magnitude:
\begin{eqnarray}
	p_{\mathrm{low}}(a) &=& 2 (1 - a)  \nonumber \\
    p_{\mathrm{flat}}(a) &=& 1 \nonumber \\
    p_{\mathrm{high}}(a) &=& 2 a,
\end{eqnarray}
where $a = |c\mathbf{S}/(Gm^2)|$ is the spin magnitude, ranging from zero to one.  
In addition, we consider both isotropic and aligned spin models~\citep{vlvrs:2014, Vitale:2015tea, sbm:2017, Farr_2017, thrane_talbot:2018}.  In the isotropic model each of the component spins is isotropically distributed on the sphere. In the aligned spin model, the spin is assumed to be  aligned with the orbital angular momentum, i.e. $\chi \equiv a$. We assume a specific distribution of all the parameters, varying only the distribution of the effective spin, $\chi_{\mathrm{eff}}$ (as defined in Eqn.~\ref{eq:chi_eff}) between the models and perform model selection between these to determine which of the models are preferred by the observations to date. The figures in this paper use acronymns for model names based on their initials i.e. isotropic models have initial (I) and aligned models have initial (A).

\subsection{Two Primary Effects}
Before we proceed to model selection we must consider two primary effects black hole spins have on the waveform. 

First, we must note the degeneracy between measurements of the effective spin parameter $\chi_{\mathrm{eff}}$ and the mass ratio $q = m_2/m_1$ of the binary (where we require $m_{2} \le m_{1}$ so that $q \le 1$). A significant source of measurement error in $\chi_{\mathrm{eff}}$ is due to the  partial degeneracy between mass-ratio and effective spin. While the two black holes are in a slowly inspiralling orbit, the emitted waveform is well approximated by the post-Newtonian approximation.  There, the dominant term that affects the  phase evolution of a binary is its chirp mass, $\mathcal{M} =  m_1^{3/5} m_2^{3/5} (m_1 + m_2)^{-1/5}$.  The following terms are dependent upon the binary's mass-ratio and the black hole spins \citep{Cutler:1994ys,Poisson:1995ef,baird-2013}. At this order, a high $\chi_{\mathrm{eff}}$ - low $q$ binary is indistinguishable from a low $\chi_{\mathrm{eff}}$ - high $q$ binary.  Consequently, for lower mass black hole binaries, where the majority of the gravitational wave signal is observed in the inspiral, the effective spin and mass ratio will be degenerate. For higher mass binaries, only the last few orbits of the binary are observed and the merger and ringdown of the binary provides a significant contribution to the observed signal.  The merger and ringdown parts of the waveform depend primarily on the total mass of the system. Thus, for higher mass binaries, the degeneracy between mass ratio and spin is less pronounced \citep{hwb:2016}. The spin distribution can be obtained by marginalizing over the joint estimate of the mass and spin distribution $p(m_{1}, m_{2}, \chi_{1}, \chi_{2})$. Thus, given the degeneracies between measurements of spins and masses, the assumed mass distribution of binaries in the universe will have a significant impact upon the inferred spins. 

Secondly, there is a selection bias affecting inference of spin population. The overall amplitude of the emitted gravitational wave depends upon both the masses and effective spin of the binary. The amplitude of the gravitational wave scales with the mass of the system.  In addition, a spin-orbit coupling causes binaries with a positive $\chi_{\mathrm{eff}}$ to undergo a greater number of orbits prior to merger and emit larger amplitude gravitational waves, compared to those with a zero or negative $\chi_{\mathrm{eff}}$ \citep{Campanelli:2006}. For example, a 40 M$_{\odot} \textrm{--}$ 30 M$_{\odot}$ binary, with maximal aligned spins, can be observed to a distance 1.6 times as large as one with maximal anti-aligned spins, leading to a factor of four increase in the observed rate for a population uniformly distributed in volume. We are biased towards observing high-$\chi_{\mathrm{eff}}$ binaries, so the fact that binaries with high, aligned spins can be observed at greater distances means that these will be preferentially observed, and their  non-observation after six detections suggests that high-$\chi_{\mathrm{eff}}$ binaries are rare. Moreover, the increase in distance to which binary is observable, caused due to increase in the value of $\chi_{\mathrm{eff}}$ is also mass dependent. Heavier binaries get a greater push in distances to which they can be observed than the lighter ones. Thus, selection effects will also depend on the assumed mass distribution.

Ideally, a combined analysis of masses and spins will naturally account for these effects. A flexible non-parametric prior, that maximizes the overall probability of observing all the gravitational wave signals, can be used to obtain the parameter distributions. Such an analysis will require hundreds of events (See also \cite{Wysocki:2018mpo} which constructs a phenomenological distribution with limited number of gravitational wave observations). 

At present, we have only a limited number of observations and therefore we account for these effects by imposing an astrophysically motivated mass distribution on the universe:  $p(m_1) \propto m_1^{-2.3}$ with $m_2$ uniformly distributed between 5 M$_\odot$ and $m_1$. The choice is based on astrophysical observations that support the stellar initial mass function to follow a power-law distribution. Moreover, the power-law model provides a binary mass-ratio distribution supported by the population synthesis models \cite{dominick1:2012,rcr:2016}.  Furthermore, independent of the assumed spin distribution, the GW measurement also support power-law model ~\cite{gw170104}. In summary, power-law is among simple models that are supported by the data.

\section{Method}

Using the observed measurements of the effective spin from the six BBH mergers considered here, we use Bayesian model selection to calculate the odds ratio between the different models.  While model selection is quite  standard, care must be taken to ensure that the selection effects and mass priors are correctly incorporated; see also \cite{loredo_2005, Mandel:2016}. We are interested in calculating
\begin{equation}\label{eq:bayes}
p(\lambda | \{ \bm{d}\}) = \frac{p( \{ \bm{d} \} | \lambda) p(\lambda)}{p( \{ \bm{d} \} )} \, ,
\end{equation}
where $\{ \bm{d} \}$ denotes the set of observations, $p(\lambda)$ is the prior on the model $\lambda$ and $p( \{ \bm{d} \})$ is
formally given as the integral over $\lambda$ of the numerator:
\begin{equation}\
p(\{ \bm{d} \} ) = \int d \lambda p( \{ \bm{d} \} | \lambda) p(\lambda).
\end{equation}

Since the model, $\lambda$, gives a distribution for the parameters of the signal, $\bm{\theta}$, we can express the probability of obtaining a given data set $\bm{d}$ corresponding to a single observation as
\begin{equation}\label{eq:p_d_given_lambda}
p( \bm{d} | \lambda) = \int \bm{d} \bm{\theta} p(\bm{d} | \bm{\theta} ) p (\bm{\theta} | \lambda),
\end{equation}
where the distribution of $\bm{d}$ given $\bm{\theta}$ is calculated from the Gaussian likelihood as
\begin{equation}\label{gauss_lklhd}
p\left(\bm{d} | \theta \right) \propto \prod_{X \in \mathrm{dets}}
\exp \left[ -\frac{1}{2} \langle d_{X} - h_{X}(\bm{\theta})| d_{X} - h_{X}(\bm{\theta}) \rangle\right] \, ,
\end{equation}
where $d_{X}$ denotes the data in detector `X' and and $h_{X}(\bm{\theta})$ is the gravitational waveform expected in 
detector `X' from a binary with parameters $\bm{\theta}$. The product is over detectors in the network,
$ \langle a | b \rangle$ is the noise weighted inner product, defined in the frequency domain as,
\begin{equation}
\langle a| b\rangle  = 4 \, \mathrm{Re} \int_{0}^{f_{\mathrm{max}}} \frac{\tilde{a}(f)\tilde{b}(f)^{\star}}{S(f)} \; \D f, \label{mfilter}
\end{equation}
and $S(f)$ is the power spectrum of the detector noise \citep{Cutler:1994ys}. 

We must also take into account the fact that there is a separate threshold on the search.  This arises in the normalization of the probability density for $\bm{d}$ above.  When there is no threshold, the probability distribution in Equation \ref{eq:p_d_given_lambda} is  correctly normalized.  However, when we impose a threshold, we must take into account that not all sets of parameters  $\bm{\theta}$ are equally likely to lead to the identification of a gravitational wave signal.  Thus, to normalize the probability, we must integrate the probability over all realizations of the data, $\bm{d}$, which produces an event above the threshold $\rho_{\star}$.  
\begin{eqnarray}
  p_\mathrm{det}(\lambda) &:=& \int_{\rho(\bm{d}) >\rho_{\star}} \D \bm{d} \int \D \bm{\theta} p(\bm{d}| \bm{\theta}) p(\bm{\theta}|\lambda) \nonumber \\
  &=& \int \D \bm{\theta} \left[ \int_{\rho(\bm{d}) >\rho_{\star}} \D \bm{d}  p(\bm{d}| \bm{\theta}) \right] p(\bm{\theta}|\lambda) \nonumber \\ 
  &=& \int \D \bm{\theta} p_{\mathrm{det}}(\bm{\theta}) p(\bm{\theta}|\lambda), \label{v_vis_bayes}
\end{eqnarray}
where we have introduced the quantity $p_{\mathrm{det}}(\bm{\theta})$ that encodes the probability of an event with
parameters $\bm{\theta}$ being observed above threshold.  Thus, the normalization factor is simply the probability of
an event drawn from the population described by $\lambda$ leading to a signal in the data that is above the detection 
threshold.  

In principle $\bm{\theta}$ includes all parameters required to fully describe the system: sky location, distance, binary
orientation, masses and spins.  However, for the discussion here, it is more useful to restrict to $\bm{\theta} = (m_{1}, m_{2},
\chi_\mathrm{eff}$).  Then, the detection probability $p_{\mathrm{det}}(\bm{\theta})$ can also be expressed as the fraction 
of sources with a given set of masses and spins that are observed, where we marginalize over the other parameters.
We assess this by distributing sources uniformly in space-time volume and orientation, and evaluating the fraction
that will be observed above the detection threshold, i.e.
\begin{equation}\label{pdf_obs}
p_{\mathrm{det}}(\bm{\theta}) = \frac{1}{V_0} \int_{0}^{z_{\mathrm{max}}}dz\frac{dV_c}{dz}\frac{1}{1+z}f(z, \bm{\theta}), 
\end{equation}
where $V_c$ is the co-moving volume, $0 \leq f(z, \bm{\theta}) \leq 1$ is the selection function giving the probability of 
detecting a source with parameters $\bm{\theta}$ at redshift $z$ and the factor of $1 + z$ is due to time dilation caused 
by the expanding universe.  $V_{0}$ is the total volume within redshift $z_{\mathrm{max}}$ and serves as a normalization 
factor.  Then $p_{\mathrm{det}}(\lambda)$ is the population averaged detection probability, which is proportional to the 
sensitive volume $V_{\mathrm{pop}}(\lambda)$, defined as,
\begin{equation}
	\frac{V_{\mathrm{pop}}(\lambda)}{V_{0}} = p_{\mathrm{det}}(\lambda) = \int d \bm{\theta} p(\bm{\theta}|\lambda) 
	p_{\mathrm{det}}(\bm{\theta}),
    \label{eq:v_pop}
\end{equation}
where,
\begin{equation}
 V_0 = \int_0^{z_\mathrm{max}} \frac{\D V_c}{\D z} \frac{1}{1+z} \;\D z, \label{eq:V0}
\end{equation}
is the total volume.
Sensitive volume is a primary ingredient in accounting for the selection effects and can be estimated using 
semi-analytical~\citep{gw150914_rates} or numerical methods \citep{scale_injections_2018}.

We can now express the distribution for the data $\bm{d}$, corresponding to a single observation, given the model $\lambda$ as
\begin{equation}
p( \bm{d} | \lambda) = \frac{ \int \bm{d} \bm{\theta} p(\bm{d} | \bm{\theta} ) p (\bm{\theta} | \lambda)}{p_{\mathrm{det}}(\lambda)} \, .
\end{equation}
The expression generalizes in a straightforward manner to a population of observed events, as we assume that the
parameters of the signals are independent, such that
\begin{equation}
p( \{ \bm{d} \} | \lambda) = \prod_{i=1}^{N} p(\bm{d}_{i} | \lambda) \, .
\end{equation}
Finally, we can use Equation \ref{eq:bayes} to obtain the probability of a model $\lambda$ given the set of observations $\{ \bm{d} \}$ as:
\begin{equation}
p(\lambda | \{\bm{d}\}) =  \frac{p(\lambda)}{ \left[p_{\mathrm{det}}(\lambda) \right]^{N} p( \{ \bm{d} \} )}
\prod_{i=1}^{N}  \int d \bm{\theta} p(\bm{d}_{i} | \bm{\theta} ) p (\bm{\theta} | \lambda) \, .
\end{equation}
This can then be used in a straightforward manner to perform model selection between two models $\lambda_{1}$
and $\lambda_{2}$ as
\begin{equation}\label{eq:odds_ratio}
\frac{p(\lambda_1 | \{\bm{d}\} )}{p(\lambda_2 | \{\bm{d}\} )} = 
\left[\frac{V_{\mathrm{pop}}(\lambda_1)}{V_{\mathrm{pop}}(\lambda_2)}\right]^{-N} \prod_{i=1}^{N} 
\left[\frac{ \int \D \bm{\theta} p(\bm{d}_i | \bm{\theta}) p_{\text{pop}}(\bm{\theta} |\lambda_1) }{
\int \D \bm{\theta} p(\bm{d}_i | \bm{\theta}) p_{\text{pop}}(\bm{\theta} |\lambda_2) }\right]
\left[\frac{p(\lambda_{1})}{p(\lambda_{2})}\right]
\end{equation}
The three terms in the odds ratio are easily understood.  The final term is simply the ratio of the priors for the two models.  In this paper, when comparing models, we take an equal prior probability for the models so this term is equal to unity.  The middle term is the probability for observing the data $\bm{d}_{i}$ given the model $\lambda$ and the first term arises as a  normalization due to the threshold in the identification of signals in the data.  We note that the overall prior on the data $p(\{\bm{d}\})$ cancels as it appears in the same way for both the models.

Rather than repeating the parameter estimation with a number of different prior distributions for $\bm{\theta}$, it is more straightforward to perform it once, with a simple prior and then to re-weight the samples.  In particular, let us assume that we obtain an estimate of the parameters $p_{\mathrm{PE}}(\bm{\theta} | \bm{d})$ given a prior $\pi(\bm{\theta})$.  We can then use this to obtain an estimate of the conditional probability for $\bm{d}$:
\begin{equation}
p(\bm{d} | \bm{\theta}) = \frac{ p_{\mathrm{PE}}(\bm{\theta} | \bm{d}) p(\bm{d})}{\pi(\bm{\theta})}
\end{equation}
When performing parameter estimation, we obtain a set of posterior samples $\theta^{j}$ that describe the posterior distribution for $\bm{\theta}$.  Explicitly, we can approximate an integral over the parameter space as
\begin{equation}
\int \D \bm{\theta} \, p_{\mathrm{PE}}(\bm{\theta} | \bm{d}) f(\bm{\theta}) \approx \sum_{j} f(\theta^{j}) \, .
\end{equation}
Thus, the integral in Equation \ref{eq:odds_ratio} can be well approximated by a sum over the (appropriately weighted) posterior samples
\begin{equation}\label{eq:final_odds}
\frac{p(\lambda_1 | \{\bm{d}\} )}{p(\lambda_2 | \{\bm{d}\} )} \approx
\left[\frac{V_{\mathrm{pop}}(\lambda_1)}{V_{\mathrm{pop}}(\lambda_2)}\right]^{-N} \prod_{i=1}^{N} 
\left[\frac{ \sum_{j} p(\theta_{i}^{j} | \lambda_{1}) / \pi(\theta_{i}^{j})}{ \sum_{j} p(\theta_{i}^{j} | \lambda_{2}) / \pi(\theta_{i}^{j})} \,
\right]
\left[\frac{p(\lambda_{1})}{p(\lambda_{2})}\right].
\end{equation}
The first term gives the ratio of the sensitive volumes for the two models, and favours the model with the lower sensitive volume.  The second term sums over the re-weighted posterior samples, where the re-weighting factor is simply the ratio of the desired prior to the one used in the parameter estimation. The final term is the ratio of the priors of the two models.  Let us now look in detail at the impact of the three factors when 
calculating the odds ratios.  As discussed previously, we will always assume an equal prior between models, so the final term is unity.

In order to estimate the sensitive volumes of the different population models, defined in Equation \ref{eq:v_pop}, we perform Monte Carlo integration. To do so, we sample from the astrophysically expected distributions of parameters and determine the fraction of sources that would be observed.  The six populations used in the analysis follow the same mass distribution but different spin distributions.  Random samples of the masses and spins are drawn from the population and are assigned 
randomly chosen orientation and sky location. Samples are distributed in redshift as determined by standard cosmology. The expected signal-to-noise ratio (SNR) of the signals produced by these binaries at the detectors are calculated and signals that cross a certain SNR threshold are labeled as recovered.  Since all of the events apart from GW170814 were observed by only the LIGO detectors, for simplicity we estimate the sensitive volume for the LIGO Hanford -- LIGO Livingston detector network.
We choose a fixed power spectrum for the detector noise operating close to the sensitivity of the LIGO detectors during the first observing run.  While the sensitivity of the detectors varies over the runs, and between the first and second observing runs, the ratio of sensitive volumes for the different population models is relatively insensitive to changes in the detector sensitivity.  To optimize the calculation, we estimate the SNR for face-on signals on a fiducial grid of binary masses and spins.  The expected SNR of a binary with arbitrary masses, spins, location and orientation is calculated 
by linear interpolation in the mass and spin space and incorporating the loss in SNR due to the arbitrary orientation \cite{schutz:2011}. To test the efficacy of the procedure, we compare our result with the results reported in reference \citep{scale_injections_2018}. Our results are within 10\% of the reported values.

The Monte Carlo equivalent of Equation \ref{eq:v_pop} is given by
\begin{equation}
 V_{\mathrm{pop}}(\lambda) = V_0 \frac{\mathrm{N}_{\mathrm{rec}}}{\mathrm{N}},  \label{eq:vol_mc}
\end{equation}
where $\mathrm{N}$ is the number of samples used in simulating the population and $\mathrm{N}_{\mathrm{rec}}$ is the number of recovered samples.

\begin{figure}
\centering
\includegraphics[width=.9 \linewidth]{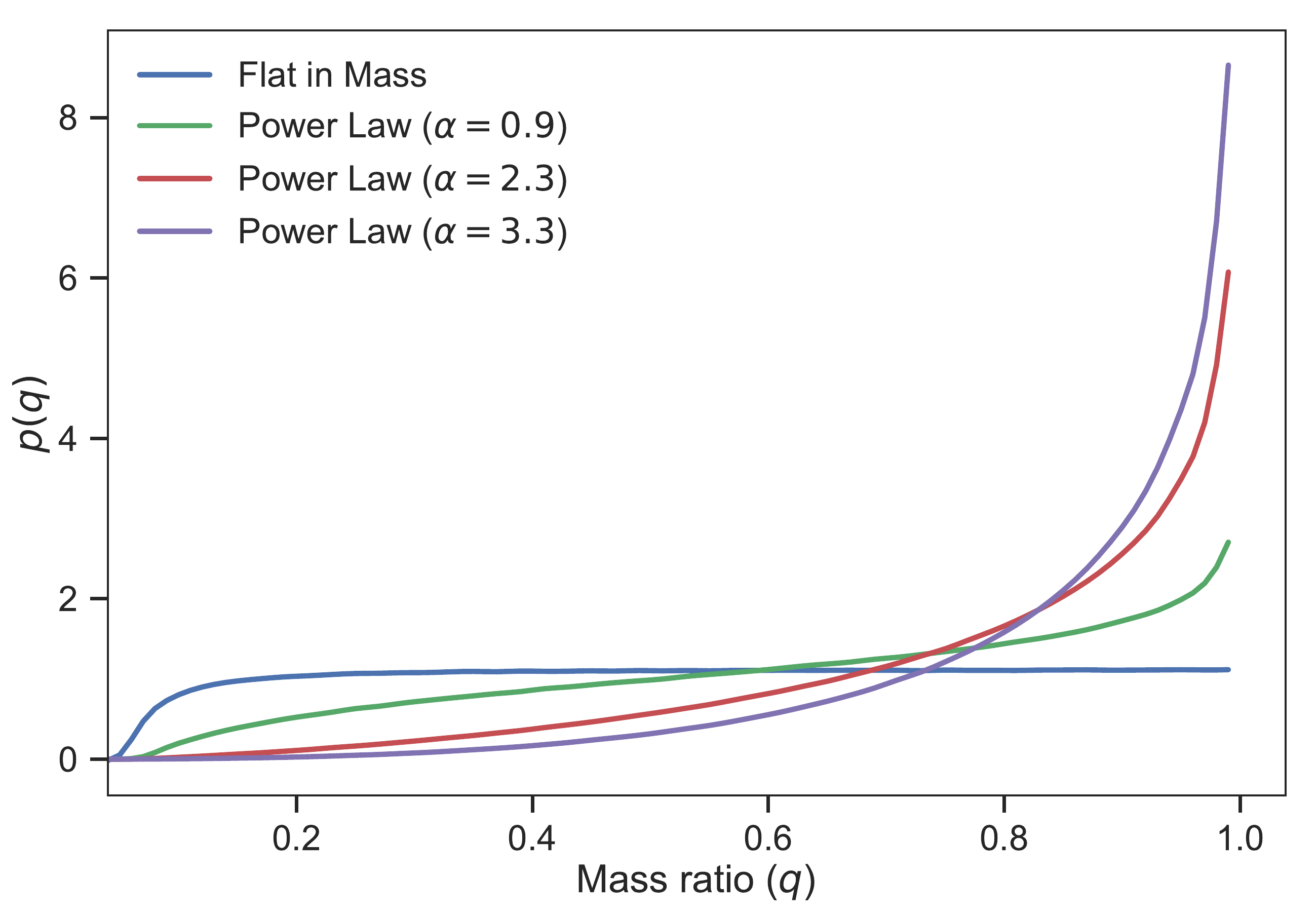}
\caption{Distribution of mass ratio for various mass distributions.  We select $p(m_{1}) \propto m_{1}^{-\alpha}$ and $p(m_{2})$ uniform in $m_{2}$ between 5 M$_{\odot}$ and $m_{1}$.  The mass ratio distribution for different  values of the power law index $\alpha = 0.9, 2.3$ and $3.3$ is shown.  At larger values of $\alpha$, binaries of comparable mass are favoured.  We also show the mass ratio distribution from a flat in $m_{1}$ and $m_{2}$ distribution.   }
\label{fig:mass_ratios}
\end{figure}

Next, let us consider the re-weighting of the posterior samples.  In particular, the prior, $\pi(\bm{\theta})$ is usually taken to be flat in $m_{1}$ and $m_{2}$, subject to the condition that $m_{1} > m_{2}$ and flat in the z-components of the spin \citep{first_monday_pe, O1:BBH, gw170104}.  In computing the probabilities for the various  models under consideration we must vary the spin prior to match one of the six distributions under consideration.  In addition, we would like to use a different mass prior, which is astrophysically motivated~\citep{fishbach_holz:2017,Talbot:2018cva}.  In particular, we select $p(m_{1}) \propto m_{1}^{-\alpha}$ and $p(m_{2})$ uniform in $m_{2}$ between 5 M$_{\odot}$ and $m_{1}$ \citep{gw150914_rates, O1:BBH}. In Figure \ref{fig:mass_ratios} we show the  prior distribution for the mass ratio, given the flat prior $\pi(\bm{\theta})$.  In addition, we show the distribution that is obtained with the power law prior, with $\alpha = 2.3$, the value used to obtain the results, as well as $\alpha = 0.9$ and $\alpha = 3.3$. These values, with mean at $\alpha = 2.3$, cover one-sigma confidence interval of the possible values of $\alpha$ that are consistent with the observations \citep{gw170104_supplement}.  

The mass-ratio distribution based on an astrophysical model is significantly different from the one obtained with a flat prior on the component masses.  In particular, close to equal mass ratio systems are significantly more likely while binaries with mass ratios greater than 5:1 are down-weighted by a factor between 1.5 and 30, depending upon the value of $\alpha$.  The higher the value of the power-law  slope $\alpha$, the more the distribution is skewed towards equal masses.  Since there is a degeneracy between mass-ratio and aligned spin, a preference for close to equal mass binaries will provide tighter posteriors on the spin. So, applying an astrophysically motivated mass prior leads to a preference for lower spin values.  This has a significant impact in down-weighting the high-spin distributions when summing over the re-weighted posterior distributions in Equation \ref{eq:final_odds}.  

Finally, we discuss the generation of the posterior samples.  The full LIGO-Virgo analysis produces thousands of posterior samples which are used to obtain the parameter distributions presented in the results papers.  In principle, those could be used directly in the calculation of the odds ratio in Equation \ref{eq:final_odds}.  However, at present only samples corresponding to LIGO's first scientific run are publicly available \citep{losc} so we instead generate distributions that mimic the full parameter estimation results, based upon the parameter values and uncertainties presented in the results papers.  As we are interested in only the masses and spins, we do not perform a full parameter estimation analysis, but rather use the fact that the measurement of masses  and spins is largely independent of the sky location and binary orientation \citep{bayestar}.  

We approximate the detector data $\bm{d}$ using a gravitational waveform \citep{phenomd1, phenomd2} corresponding to the median value of the reported values of the masses, both aligned spin magnitude for both black holes set equal to the reported median value of $\chi_{\mathrm{eff}}$ as reported in the observations \citep{O1:BBH, gw170104, GW170608, GW170814}, and zero precessing component of the spins.  In addition, we generate the waveform associated with a face-on binary (which is consistent with all of the observed signals), and at a distance $D$ that produces an SNR equal to the reported value.  To obtain parameter estimates, we then draw a large number of samples in mass, spin and distance, with the  distributions following the prior distribution $\pi(\bm{\theta})$.   Specifically, we use flat distributions of component masses (with limits between $5 M_{\odot}$ and $75 M_{\odot}$), dimensionless aligned spin magnitudes (between 0 and 0.95),  and distances distributed uniformly in volume (between $D/2$ and $2D$).  For each of these points, we generate a face-on signal with the appropriate masses, spins and distance and calculate the Gaussian likelihood defined in Equation \ref{gauss_lklhd}.  We are only interested in mapping the degeneracy between mass ratio and spin and as we assign the same noise to the detectors in the network, the properties of the detector network are not important and we perform the calculation using a single detector, with a sensitivity  matching that obtained by the LIGO detectors during O1. Finally, we estimate the posteriors on the parameters by performing rejection sampling based on the  value of the calculated likelihood. By assuming face-on signals, we will obtain a more tightly constrained distance distribution than when allowing arbitrary orientations.  While our calculations do not directly use the distance,  the distribution will impact the inferred mass distributions as the signal observed at the detector depends upon the redshifted masses $m_{i} (1 + z)$.  However, in this work, we are primarily concerned with the mass-ratio and spin of the binary, both of which are unaffected by redshift effects.  Thus, this simplified method of obtaining posterior samples will provide accurate distributions for the most important parameters in this analysis.

Table \ref{tab:compare_interval} compares the credible intervals for the announced GW observations 
\citep{O1:BBH, gw170104, GW170608, GW170814} with the credible intervals of the samples obtained using the method  described above. The intervals obtained for the masses and effective spins are comparable. We have made several  approximations which we would  expect to give differences at the observed levels.  In addition, the reported gravitational wave results make use the average results from both spin-aligned and full-spin precessing  models, while in the posteriors that we generate we consider only aligned spins. The mass-ratios of the approximated posteriors  are nearly equal to the reported values. We have verified that our results are insensitive to shifts in the posterior distributions at the levels reported in Table \ref{tab:compare_interval}.

\begin{deluxetable*}{||c c c c||}
\tablecaption{Comparison between reported credible intervals of gravitational wave observations and those obtained from the approximate posterior samples generated using the methods discussed in the paper.  In each entry in the table, the first number and uncertainty gives the median and 90\% range obtained from the posterior samples generated for our analysis.  The second number and interval give the same values as derived directly from the gravitational wave data  \citep{O1:BBH, gw170104, GW170608, GW170814}.  The median values are all in good agreement (less than $1\sigma$ deviation
in all cases), and the intervals are also comparable. \label{tab:compare_interval}}
\tablehead{
  & Primary Mass & Secondary Mass & Effective Spin\\ 
  & $m_1^{\mathrm{source}}/\mathrm{M}_\odot$ & $m_2^{\mathrm{source}}/\mathrm{M}_\odot$ & $\chi_{\mathrm{eff}}$} 
\startdata
 GW150914 & $36.0_{-5.0}^{+6.0}\;/\;36.2_{-3.8}^{+5.2}$ & $27.0_{-4.0}^{+4.0}\;/\; 29.1_{-4.4}^{+3.7}$ & $-0.05_{-0.12}^{+0.11}\;/\;-0.06_{-0.14}^{+0.14}$ \\ 
 \hline
 LVT151012 & $24_{-6}^{11}\;/\;23_{-6}^{+18}$ & $13_{-4}^{+4}\;/\;13 _{-5}^{+4}$ & $0.1_{-0.2}^{+0.2}\;/\;0.0_{-0.2}^{+0.3}$ \\
 \hline
 GW151226 & $15.0_{-4.0}^{+12.0}\;/\;14.2_{-3.7}^{+8.3}$ & $7.0_{-2.0}^{+3.0}\;/\;7.5_{-2.3}^{+2.3}$ & $0.22_{-0.13}^{+0.29}\;/\;0.2_{-0.1}^{+0.21}$ \\
 \hline
 GW170104 & $29.0_{-4.0}^{+9.0}\;/\;31.2_{-6}^{+8.4}$ & $20.0_{-4.0}^{+4.0}\;/\;19.4_{-5.9}^{+5.3}$ & $-0.09_{-0.18}^{+0.13}\;/\;-0.12_{-0.2}^{+0.21}$ \\
 \hline
 GW170608 & $13_{-3}^{+12}\;/\;12_{-2}^{+7}$ & $7_{-3}^{+2}\;/\;7_{-2}^{+2}$ & $0.13_{-0.12}^{+0.34}\;/\;0.07_{-0.09}^{+0.23}$ \\
 \hline
 GW170814 & $30.0_{-4.0}^{7.0}\;/\;30.5_{-3.0}^{+5.7}$ & $23.0_{-4.0}^{+3.0}\;/\;25.3_{-4.2}^{+2.8}$ & $0.07_{-0.13}^{+0.13}\;/\;0.06_{-0.12}^{+0.12}$ \\ [1ex] 
\enddata
\end{deluxetable*}

\section{Results and Discussion} \label{sec:results}

\begin{figure*}[ht!]
\plotone{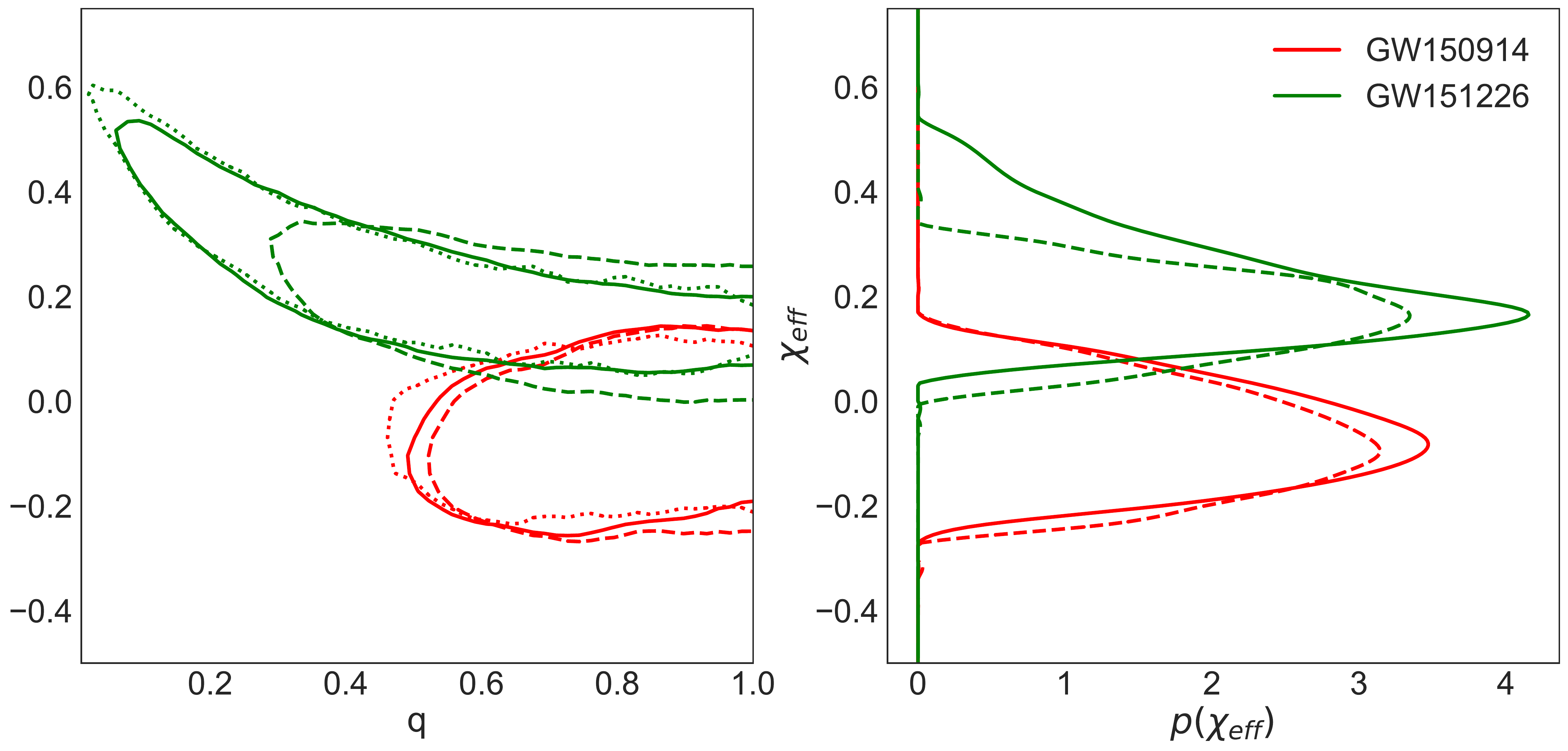}
\caption{Contour plots for the posterior samples on the $q\textrm{--}\chi_{\mathrm{eff}}$ plane. Contours show the 90\% credible regions  of the posterior probability. The solid line corresponds to the posterior probability estimated using publicly available posterior \citep{losc}. Samples have been produced using a flat prior distribution for the component masses. The dashed line corresponds to the posterior probability obtained by re-weighting the samples with a power law distribution for $m_1$: $p(m_1) \propto m_1^{-2.3}$, and $m_2$ uniformly distributed  between 5 M$_\odot$ and $m_1$. The posterior probability on the effective-spin is affected by the choice of prior on the mass-ratio. Power-law prior disfavours small mass-ratio, which restricts the inferred spin distribution. For comparison, plot also includes dotted contours prepared using mock samples that are used in this analyses. The true and mock samples are in good agreement.} 
\label{fig:resampled_chieff}
\end{figure*}

We now investigate the impact of both mass priors and selection effects on the inferred spin distribution for black holes  in binaries. Some of the previous work have used $\chi_{\rm eff}$ in inferring spin distributions~\citep{Farr_2017, farrb2018}, however, inclusion of these two effects allows tighter constraints on the fraction of binaries with large values of $\chi_{\mathrm{eff}}$.  The effect of using an astrophysically motivated mass prior can be seen in Figure \ref{fig:resampled_chieff}, that shows the mass ratio and effective spin contours for GW150914 and the much lighter GW151226.  As expected, the low mass systems display significant correlations, which are less evident at higher masses.  The figure shows the inferred mass ratio and effective spin distributions for two choices of prior: a flat-in-mass prior and the astrophysically motivated power law.  The flat-in-mass distribution allows for more extreme mass ratios, and consequently larger values of the effective spin. For example, the flat prior on masses, which was used in producing parameter estimates in the LIGO-Virgo results papers, follows spins up to $0.6$, while imposing the astrophysically motivated mass distribution removes support for $\chi_{\mathrm{eff}} > 0.4$ for GW151226. However, for high-mass systems, such as GW150914, where there is less degeneracy between masses and spins, the effect is not significant.%
\footnote{A similar analysis of the effect of mass priors has been performed in \citep{Vitale_2017}.}. Figure \ref{fig:resampled_chieff} also plots contours produced from publicly available parameter samples. Considering also Table \ref{tab:compare_interval}, mock samples show good agreement with the true samples.

\begin{figure}
\centering
\includegraphics[width=.9 \linewidth]{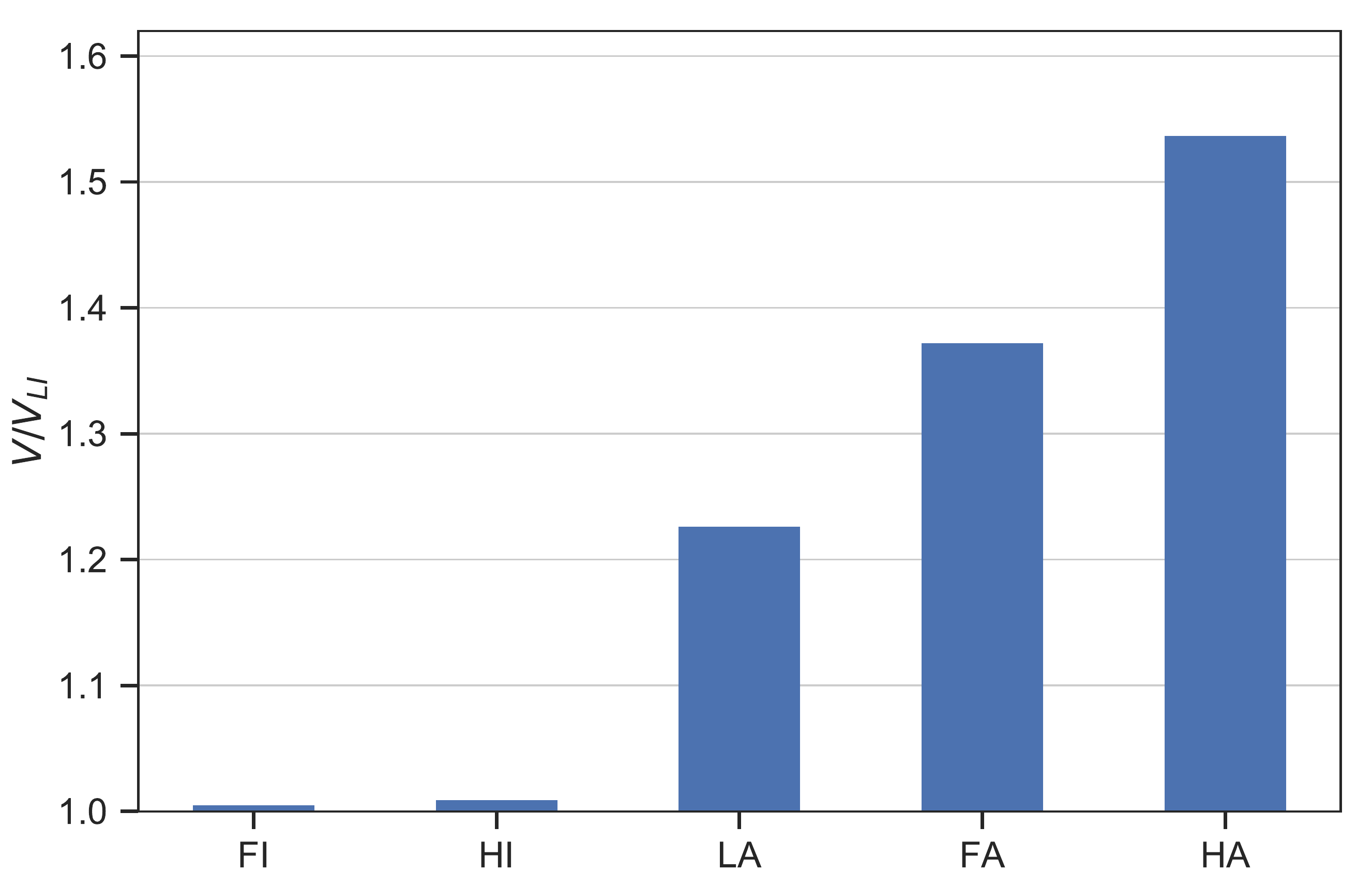}
\caption{Sensitive volumes of the spin models compared to the low isotropic model. The figure plots the ratio of sensitive volumes of various models with respect to the low isotropic spin model. This ratio raised to the power by the number of observation accounts for the selection effects.}
\label{fig:vols_spin_mod}
\end{figure}

Impact of selection effects are shown in Figure \ref{fig:vols_spin_mod} that plots the ratio of the sensitive volume of the spin models and the sensitive volume of the  low isotropic model. Binaries with a higher spin magnitude can be observed at a greater distance,  so we expect that the low isotropic model, which leads to the population with the smallest spin magnitudes, to have the lowest sensitive volume.  Thus, all other things being equal, for each event observed the model with the lower sensitive volume is preferred.  This has a significant impact for the aligned spin models, but the volume ratio between models with isotropic spin distributions is close to unity.  Thus, this doesn't have a significant impact for a small number of events, but does give a factor of 5 contribution to the odds ratio with 50 events.

\begin{figure}
\centering
\includegraphics[width=.9 \linewidth]{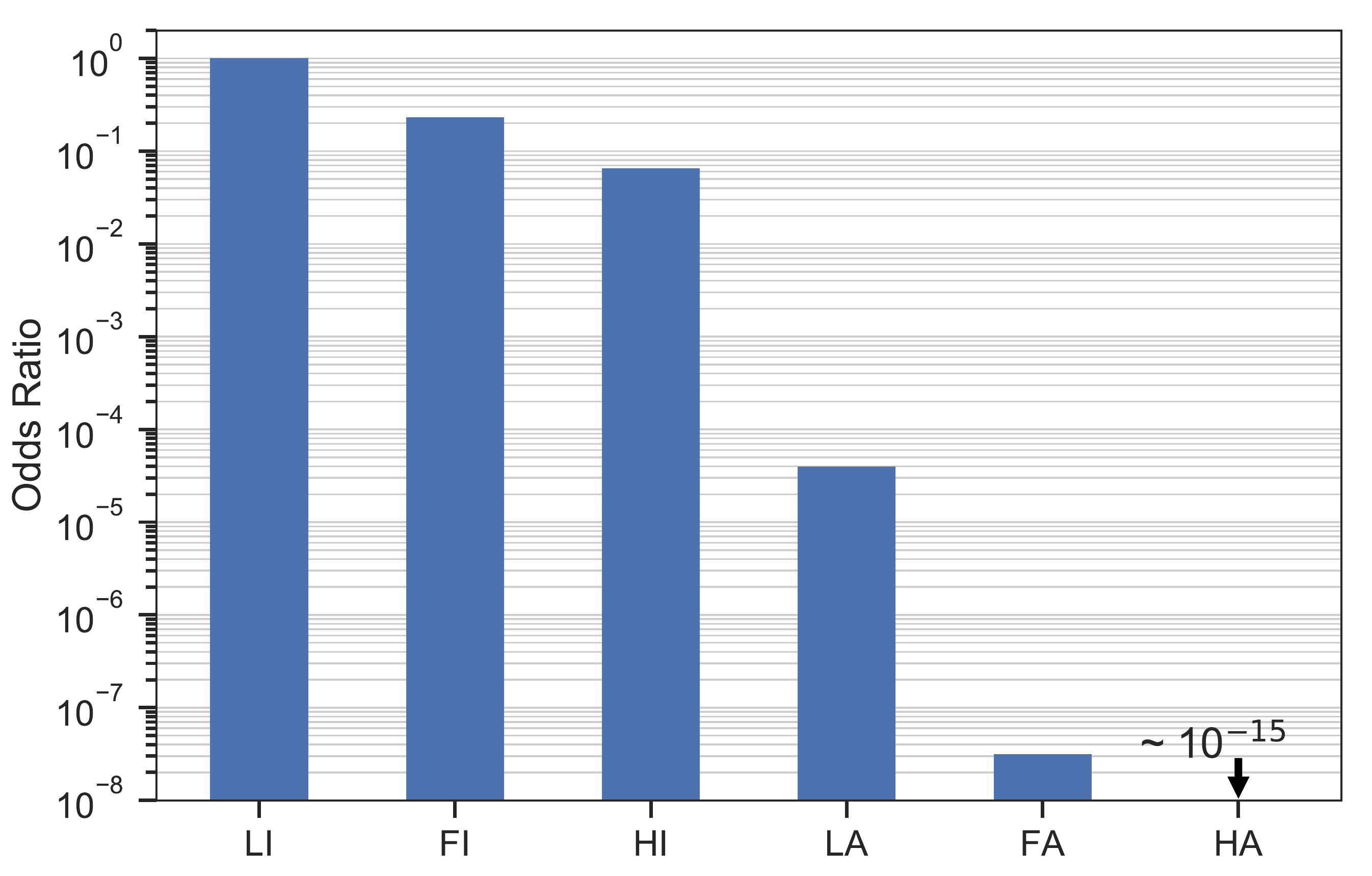}
\caption{Odds ratios for different spin magnitude and orientation distributions in reference to low isotropic model. All of the aligned spin distributions are excluded at $>$4$\sigma$; the high-isotropic distribution is disfavoured by a factor of 15:1, 
which equates to about $1.8 \sigma$.}
\label{fig:results}
\end{figure}

The results of the analysis are shown in Figure \ref{fig:results}. This shows the odds ratio between each of the six models discussed above and the low isotropic model.  The low isotropic distribution is preferred.  All models with aligned spins are disfavoured at greater than \alignedodds  or, equivalently,  \alignedsigma.  (The flat- and high-aligned distributions are disfavoured at $>$5$\sigma$.)  This provides strong evidence against aligned spins.  The improvement arise from three different factors.  The inclusion of two additional events (GW170608 and GW170814) increases the odds ratio by a factor of six, while an accurate treatment of the selection effects and the use of an approximately good mass ratio distribution to handle the mass ratio--spin degeneracy increase the odds by a factor of sixty.  Incorporation of selection effects increases the odds ratio by a factor of four while accounting for correlations between mass ratio and spin increases the odds ratio by a factor of sixteen. Thus, these three effects combined explain the factor of around four hundred improvement in our ability to exclude aligned spin models in favour of isotropic distribution of spins. 

Although LA model does not represent a viable astrophysical model for the spin distribution the systematic and selection effect are most visible for this model. Standard posterior samples of both GW151226 and GW170608 show evidence of alignment and have an odds-ratio of greater than one for LA spin model compared to LI model. Observations are best described by a mixture of LA and LI spin models and a spin analysis aimed at mixture models is best suited. Another important factor to consider is misalignment between the orbital angular momentum vector and total angular momentum vector. We note that allowing spin misalignments at merger of up to $30^{\circ}$ has only a minimal effect on the probability distribution of $\chi_{\rm eff}$.  The overall magnitude of $\chi_{\mathrm{eff}}$ will reduce by few percent due to this misalignment. There are rigorous analysis that deal with misalignment and mixture models and readers are referred to them \citep{Stevenson:2017dlk, thrane_talbot:2018}. 

However, as noted in this paper, systematic and selection effects will affect the inference made by a mixture model as it will likely introduce a bias towards LA models e.g. the odds-ratio for GW151226 reduces for the LA model on using re-weighted posterior samples instead of the standard posterior samples and odds-ratio for GW170608 reduces to close to unity.

The result is only mildly dependent upon the value of the slope, $\alpha$.  As we increase the value of $\alpha$, we  favour lower mass black hole binaries in the population while lower values of $\alpha$ have a larger fraction of high mass binaries.  The impact of spin on visible volume is more significant for higher masses so, consequently, small values of $\alpha$ lead to larger difference in sensitive volumes between different spin distributions. However, for larger $\alpha$, the mass ratio and aligned spin are more tightly constrained.  This increases the importance of re-weighting the posterior samples.  Thus, overall, changing the value of $\alpha$ has a limited impact on the results.  We also note that choice of the minimum and the maximum masses in the power-law model has only mild effect on the results. Table \ref{tab:various_bf} lists odds-ratios of HI and LA spin models in reference to LI model for different values of $\alpha$ and the maximum mass of the primary component of the binary.
 
 \begin{deluxetable*}{||c c c c||}
\tablecaption{Estimated odds-ratio for different values of $\alpha$ and maximum mass. Only the odds-ratios between LI versus HI, and LI versus LA models are included. \label{tab:various_bf}}
\tablehead{Spin Model & $\alpha = 0.9 $ & $\alpha = 2.3 $ & $\alpha = 3.3 $} 
\startdata
 High Isotropic (Maximum Primary Mass = 75 M$_\odot$) & 14.3:1 & 15:1 & 15.5:1\\ 
 \hline
 Low Aligned (Maximum Primary Mass = 75 M$_\odot$) & 14000:1 & 22000:1 & 20000:1\\
  \hline
 High Isotropic (Maximum Primary Mass = 95 M$_\odot$) & 14.8:1 & 16.1:1 & 16.2:1\\
  \hline
 Low Aligned (Maximum Primary Mass = 95 M$_\odot$) & 15900:1 & 25000:1 & 30000:1\\ [1ex]
\enddata
\end{deluxetable*}

More significantly, we are able to show that low spins are preferred to high spins, even when  isotropically distributed. Specifically, the low-isotropic distribution is preferred over the high-isotropic distribution at \highspinodds  (around \highspinsigma).  While highly spinning components can produce a binary with minimal $\chi_{\mathrm{eff}}$,  this requires a detailed balance between the two spins of the black holes --- either the positive spin from one black hole must  cancel with the negative spin of the second or the majority of the spin must be in the plane of the orbit.  Thus, the fact that $\chi_{\mathrm{eff}}$ was small for all of the first six black-hole-binary mergers  provides evidence that highly spinning stellar-mass black holes are rare. As discussed in detail in the methods section, this result \emph{does not} depend strongly on the choice of astrophysical mass distribution. 

We have shown that the first six observations of black hole binary mergers can be used to place a limit on the magnitude of  black hole spins based on gravitational wave observations.  The data show strong evidence for isotropic, rather than aligned,  spins.  Furthermore, there is emerging evidence that small low spin magnitudes are preferred to high spin magnitudes (see Figure 8 of \citep{farrb2018} as well as of \citep{Wysocki:2018mpo}).  This contrasts the nominal spin magnitudes inferred from x-ray binary observations, which is more consistent with high spins. 

We emphasise that these conclusions depend on our choices of possible spin distributions. If, for example, our low-spin distribution was restricted to much lower values of $\chi_{\rm eff}$, then aligned-spin configurations would not be so strongly disfavoured. However, this would only \emph{strengthen} our main conclusion, which is the preference for low spin magnitudes. 

Distributions of black hole spins will be further refined through future gravitational wave observations.  In the third advanced LIGO-Virgo observing run, there is an expectation of observing tens of black hole binary  mergers.  To get a sense of what we might expect, we simulated 30 observations from the low isotropic spin distribution and combined them with the six observations discussed above.  With this set of observations, we would be able to exclude a population of black hole binaries where 20\% have aligned spins and 80\% with isotropic spins with a confidence of at least $4 \sigma$.
Furthermore, we would obtain an odds ratio in favour of low isotropic  spins over high isotropic spins of around \checkme{130,000:1} ($4.5 \sigma$) and  flat isotropic of around \checkme{300:1} ($3.0 \sigma$).  Finally, we note that we  have not made use of the precessing component of the spin.  A clear observation of precession will give irrefutable evidence of spin misalignment, while observations of $\chi_{p}$ consistent with zero will provide further evidence against high spin magnitudes.

\section*{Acknowledgments}
We would thank the following for interesting and useful discussions: Will Farr, Frank Ohme, Richard O'Shaughnessy, Simon Stevenson, Eric Thrane and Vivien Raymond.  The authors were funded by the Science and Technology Facilities Council (STFC) grants ST/L000962/1 and ST/N005430/1.  MDH and SF was supported by the European Research Council Consolidator Grant 647839.


\bibliography{vaibhav}
\end{document}